\documentclass[11pt,a4paper]{article}
\pdfoutput=1
\usepackage{bm}
\usepackage{amsmath,amssymb} 

\usepackage{cite}

\usepackage{graphicx}

\usepackage{hyperref} 
\hypersetup{colorlinks=true, citecolor=blue, filecolor=black, linkcolor=blue, urlcolor=blue, pdfpagemode=UseNone}

\newcommand\ie{\textit{i.e.}\ }
\newcommand\eg{\textit{e.g.}\ }
\newcommand\cf{\textit{cf.}\ }

\newcommand{\viz}{{\it viz.}\ }
\newcommand{\half}{\frac{1}{2}}

\newcommand{\be}{\begin{equation}}
\newcommand{\ee}{\end{equation}}
\newcommand{\bea}{\begin{eqnarray}}
\newcommand{\eea}{\end{eqnarray}}
\newcommand{\TRM}[1]{#1}
\newcommand{\p}[1]{#1_{phys}}
\newcommand\re[1]{\,{\rm e}^{#1}\,}

\begin{document}
\begin{titlepage}
\begin{flushright}
\end{flushright}

\begin{center}
{\huge \bf The fate of non-polynomial interactions in scalar field theory}

\end{center}
\vskip1cm


\begin{center}
{\bf I. Hamzaan Bridle \& Tim R. Morris}
\end{center}

\begin{center}
{\it STAG Research Centre \& Department of Physics and Astronomy,\\  University of Southampton,
Highfield, Southampton, SO17 1BJ, U.K.}\\
\vspace*{0.3cm}
{\tt  I.H.Bridle@soton.ac.uk, T.R.Morris@soton.ac.uk}
\end{center}


%

\abstract{We present an exact RG (renormalization group) analysis of $O(N)$-invariant scalar field theory about the Gaussian fixed point. We prove a series of statements that taken together show that the non-polynomial eigen-perturbations found in the LPA (local potential approximation) at the linearised level, do not lead to new interactions, {\it i.e.} enlarge the universality class, neither in the LPA or treated exactly. Non-perturbatively, their RG flow does not emanate from the fixed point. For the equivalent Wilsonian effective action they can be re-expressed in terms of the usual couplings to polynomial interactions, which can furthermore be tuned to be as small as desired for all finite RG time. For the infrared cutoff Legendre effective action, this can also be done for the infrared evolution.  We explain why this is nevertheless consistent with the fact that the large field behaviour is fixed by these perturbations.
}



\end{titlepage}

\tableofcontents


\section{Introduction}\label{sec:introduction}

The Wilsonian  RG (renormalization group), in particular when adapted to the continuum (where it was christened the ``exact RG'' by Wilson \cite{Wilson:1973}), is an important framework for understanding quantum field theory outside the realm where perturbation theory is a secure guide (for introductions and reviews see refs. \cite{Morris:1998,Aoki:2000wm,Bagnuls:2000,Berges:2000ew,Polonyi:2001se,Pawlowski:2005xe,Kopietz:2010zz,Rosten:2010vm}). Without the security of perturbation theory, it is clearly important if possible to make rigorous conclusions  about such non-perturbative RG properties. A persuasive example is provided by the investigations into asymptotic safety in quantum gravity \cite{Weinberg:1980}.  Following ref.  \cite{Reuter:1996}, this has been a major area where these ideas have been applied  (for reviews and introductions see \cite{Reuter:2012,Percacci:2011fr,Niedermaier:2006wt,Nagy:2012ef,Litim:2011cp}). However this is also an area where there is little guidance from current experimental observation  or other techniques, and therefore one must place particular reliance on a rigorous understanding of the mathematical structure that the exact RG exposes, in so far as this is possible. This is especially so with recent work on ``functional truncations'' \cite{Machado:2007,Reuter:2008qx,Codello:2008,Manrique:2009uh,Benedetti:2012,Demmel:2012ub,DietzMorris:2013-1,DietzMorris:2013-2,Demmel:2014fk,Falls:2014tra,Demmel:2014hla,Dietz:2015owa,Demmel2015a,Demmel2015b,Ohta:2015efa,Percacci:2015wwa,Eichhorn:2015bna,Ohta2016,Labus:2016lkh,Dietz:2016gzg}.

In particular an analysis of what is legitimate for relevant eigen-perturbations in functional truncations \cite{DietzMorris:2013-1,DietzMorris:2013-2,Bridle:2013sra, Dietz:2016gzg},
thus determining what is the set of renormalised couplings in the continuum theory (see \eg \cite{Wilson:1973, Morris:1998}), is clearly crucial. But this has to begin by building on a thorough understanding of what is legitimate in a well-understood situation such as that of scalar field theory. Moreover, recent work has shown that, despite the complications induced by diffeomorphism invariance and background independence \cite{Reuter:2008qx,Manrique:2009uh, Bridle:2013sra}, the overall scalar factor  part of the metric (at least within the so-called conformal truncation) has RG properties that are just those of a scalar field theory -- but  crucially with wrong sign kinetic term \cite{Dietz:2015owa,Dietz:2016gzg}. Furthermore this one sign profoundly alters the RG properties (unless the field is continued to the imaginary axis) \cite{Dietz:2016gzg}. 


Clearly then it is important to begin by being sure of the facts for standard scalar field theory (\ie with right sign kinetic term). Indeed the RG properties of (standard) scalar field theory could be regarded as long settled, and in particular the classification of relevant perturbations in functional truncations \cite{MorrisReb,Morris:1996xq,Morris:1998}, see also \cite{Morris:1997xj,DAttanasio:1997he,Morris:2005ck}. However, even for perturbations around the Gaussian fixed point, a debate has continued \cite{MorrisReb,Morris:1996xq,Percacci:2003jz,Periwal:1995hw,Halpern:1997gn,Branchina:2000jp,Bonanno:2000sy,Gies:2000xr,Altschul:2004yq,Altschul:2004gt,Altschul:2005mu,Gies:2009hq,Huang:2010qn,Huang:2011xg,Huang:2011xha,Pietrykowski:2012nc,Huang:2013zaa}
over the significance of certain non-polynomial scaling solutions  to the linearised LPA (Local Potential Approximation) flow equations (\ie eigen-perturbations)
starting with Halpern and Huang's observations \cite{HHOrig,HHReply,HH2ndpaper}. Although these ideas were criticised early on \cite{MorrisReb,Morris:1996xq,Percacci:2003jz},  researchers have continued to pursue these ideas  \cite{Periwal:1995hw,Halpern:1997gn,Branchina:2000jp,Bonanno:2000sy,Gies:2000xr,Altschul:2004yq,Altschul:2004gt,Altschul:2005mu,Gies:2009hq,Huang:2010qn,Huang:2011xg,Huang:2011xha,Pietrykowski:2012nc,Huang:2013zaa}, no doubt inspired in large part by the hope that scalar fields might in certain circumstances enjoy asymptotically free interactions. This is of course of particular relevance  in four dimensions for the Higgs sector of the Standard Model, but if true would have profound consequences also for constructing theories beyond the Standard Model and indeed for continuous phase transitions in condensed matter and statistical physics. 

In this paper we re-analyse the RG properties of standard scalar field theory about the Gaussian fixed point, in particular concentrating on such non-polynomial solutions for the linearised perturbations. 
We prove a series of statements that taken together show that these solutions cannot be regarded as providing extra relevant directions. In this way we aim to consolidate as rigorously as possible what is known about scalar field theory around the Gaussian fixed point, and to settle conclusively the debate over Halpern-Huang directions. 

\TRM{Although these non-polynomial solutions were found in the LPA, and we will proceed by demonstrating that they fail to provide new relevant directions within this approximation scheme, these results extend to the derivative expansion and also to the exact equations, as we outline at the end of the paper.}


To be concrete we concentrate on the case of most interest for Higgs physics, namely
$O(N)$ invariant $N$-component scalar field theory around the Gaussian fixed point in four Euclidean space-time dimensions (where for the Higgs field, $N=4$ and the usual Wick rotation has been performed). All our arguments can however be straightforwardly adapted to other dimensions $d>2$ and to other fixed points.\footnote{In $d=2$ dimensions the situation is very different since the engineering dimension of the scalar field vanishes, leading to bounded (oscillatory) solutions for the LPA \cite{Morris:1994jc}.}
In particular in less than four dimensions,  there are non-perturbative Wilson-Fisher fixed points \cite{Wilson:1971dc,Wilson:1973}. For these fixed points, linearised solutions divide into quantised eigen-perturbations (that is scaling fields with quantised scaling dimension as is true for the polynomial solutions about the Gaussian fixed point) and non-quantised eigen-perturbations (with continuous scaling dimension as also true for non-polynomial solutions about the Gaussian fixed point) 
\cite{MorrisReb,Morris:1996xq,Morris:1998}. Since we concentrate on the Gaussian fixed point, in this paper we will use the terms (non)quantised and (non)polynomial interchangeably.

In broad outline, the arguments are as follows. The key observation 
is that solutions of the linear partial differential equation that result from linearisation of the flow equation around a fixed point, {a priori} must remain \emph{uniformly} small. That is, given an $\epsilon$ multiplying the linearised solution, there must exist a $\delta$ such that the solution is smaller\footnote{For the Gaussian fixed point `smaller' here does just mean ``less than" in magnitude. In general `smaller' is determined by ratios involving the behaviour of the fixed point itself.} than this for all values of the field. In other words it is not enough that such a perturbation is small for some given value of the field: linearisation requires that it must be small for \emph{all values} of the field. Such perturbations that grow sufficiently fast for large field violate this condition, thus ruling out the procedure used for finding them in the first place. For the quantised solutions, the mistakenly-deduced RG evolution can nevertheless be (sufficiently) recovered, while for the non-quantised solutions they cannot. Instead the large field behaviour undergoes mean-field RG evolution, which is to say in fact that in physical (unscaled) variables nothing happens: the large field behaviour remains fixed by the initial perturbation.\footnote{In fact the same is true for the quantised perturbations but for these the large field behaviour can also be interpreted as the expected RG evolution of the corresponding coupling.} Despite the fact that such a linearisation procedure has been adopted since the earliest days \cite{WegnerBook1,Wegner:1972my,Bagnuls:2000} (the so-called scaling fields), and envisioned as taking a tangent to the flow in function space, it is not a justified procedure for finding solutions that really flow into or out of the fixed point, without the uniformity condition or further analysis.

The non-quantised solutions could therefore logically be viewed as no more special than any other \emph{finite} perturbation added to the fixed point. 
Nevertheless given the significance attached to the evolution of scaling fields at finite field, we will concentrate on  their RG behaviour. We recall and then further develop detailed analysis of their renormalization group evolution, both in the UV (ultraviolet) and especially we further develop the analysis in the IR (infrared). In refs. \cite{MorrisReb,Morris:1996xq,Morris:1998} we already demonstrated that beyond a certain RG time, the IR evolution of the non-quantised perturbations lives inside the Hilbert space\footnote{\ie Banach space with inner product, not to be confused with the quantum mechanical state space.} spanned by the quantised perturbations and therefore can be expressed in terms of the IR evolution of their couplings.
We will make further progress by analysing these couplings both for the flow of the (IR cutoff) Legendre effective action \cite{Nicoll1977,Wetterich:1992,Morris:1993} and the flow of the Wilsonian effective action \cite{Wilson:1973,Polchinski:1983gv}, and in particular providing bounds on these couplings depending on $\epsilon$. In this way we are able to show that, although an infinite number of (irrelevant) quantised couplings are involved, they can be made as small as desired (for all finite
 RG time $t$) by taking $\epsilon\to0$. The RG evolution at finite field is thus expected to be
 the normal one for the universality class determined by quantised relevant couplings. 
 We explain why this is nevertheless consistent with 
the large field behaviour being still that determined by the initial perturbation.

Let us also mention that an elegant approach to studying the behaviour of the non-polynomial eigenperturbations was performed in ref. \cite{Gies:2000xr}. There the subtleties of the large $N$ limit for such perturbations, were negotiated, allowing large $N$ methods to be used to solve for the full trajectory towards the infrared ($t\to\infty$). Apart from the ``fine-tuning'' and ``naturalness'' properties found in that case, which are likely to be an artefact of $N=\infty$ \cite{Gies:2000xr}, the results are consistent with our conclusion that the RG evolution is determined at finite field by the universality class described by the quantised relevant couplings, and at large field by the original perturbation. Since the $t\to\infty$ limit is also addressed in detail, the study is complementary to ours.  On the other hand, finite $N$, the UV evolution, and the realisation that the IR evolution can be expressed in terms of the quantised couplings, were missing, and indeed do not look straightforward to access from the (somewhat subtle) $N=\infty$ limit.

In the Conclusions, sec. \ref{sec:conclusions}, we list the statements that we prove and thus this final section also provides a more detailed overview of the paper.

\section{Initial considerations}\label{sec:initial}

\subsection{Universality of  linearised LPA about the Gaussian fixed point}\label{sec:universality}

Halpern and Huang noticed that, within the (LPA) Local Potential Approximation \cite{Nicoll:1974zz,Nicoll:1976ft,Hasenfratz:1985dm,Morris:1994ki,Zumbach:1994vg,Ball:1994ji,Morris:1995af,Morris:1996kn,Comellas:1997tf} linearised perturbations about the Gaussian fixed point not only have solutions that are polynomial in the scalar field $\phi$, which can be identified with the operators usually considered in perturbation theory, but also solutions that are non-polynomial in the field with a continuous scaling dimension \cite{HHOrig,HH2ndpaper}.\footnote{Refs. \cite{HHOrig,HH2ndpaper} also proposed a line of non-perturbative fixed points following from such equations but 
this matter was conclusively settled in ref. \cite{MorrisReb} and will thus not be further discussed.} It is a particular set of these latter perturbations that Halpern and Huang identified as being asymptotically free.

They used the Wegner-Houghton equation \cite{Wegner:1972ih}. But the same equation  can also be reached by taking the sharp cutoff limit of any exact renormalisation group equation \cite{Morris:1993,Morris:1995af}. In fact after appropriate rescalings, the linearised perturbations satisfy the \emph{same} equation whatever cutoff profile is used providing it actually regulates. This is actually clear from refs. \cite{Ball:1994ji,Periwal:1995hw} for the Wilson/Polchinski version, and for the Legendre effective (average) action version
 from ref. \cite{Gies:2000xr} after finite rescalings of potential and field. The reason for this can be traced to the fact that at the linearised level, the equations are only summing up tadpole terms, whose value can be rescaled by changes of variables\TRM{ \cite{Altschul:2004yq} (see also \cite{Dietz:2016gzg}).  Indeed in this way, it is clear that the linearised flow equation  \eqref{linearised} for the potential, that we are about to derive within LPA, may in fact be derived \emph{exactly, \ie without  approximation,} whatever cutoff profile is used. We reserve further comments on this for the end of the paper.}
Thus we can easily avoid the limitations of the sharp cutoff that concerned the authors in refs. \cite{HHOrig,HHReply,HH2ndpaper}. Also we see that the non-polynomial perturbations with continuous scaling dimension are thus \emph{universal} in this sense.  

Here let us demonstrate this universality by example. At the same time we will set up the equations and notation we will need for the later discussion. We specialise to the most interesting case of four space-time dimensions (but as usual work in Wick-rotated Euclidean space).  We work with an $O(N)$ invariant effective potential $V(z,t)$, where 
\be 
\label{z}
z=\phi^a\phi_a/2 
\ee
with $\phi^a$  an $N$-component real scalar field, and 
\be 
\label{t}
t=\ln(\mu/\Lambda)
\ee 
is renormalisation group `time'.  $\Lambda$ is the running effective cutoff scale (often called $k$ in the functional renormalisation group literature) and $\mu$ is  some arbitrary fixed energy scale. We have written all quantities in dimensionless terms using the running cutoff scale, thus in terms of physical quantities 
\be
\label{dimensions}
V=\p V/\Lambda^4\qquad{\rm and}\qquad z=\p z/\Lambda^2\,.
\ee 
(The field's anomalous dimension is neglected in this version of the local potential approximation \cite{Nicoll:1974zz,Wilson:1973,Morris:1994ki}. See ref. \cite{Bervillier:2013} for discussions on this.) 
For sharp cutoff  one then has \cite{Hasenfratz:1985dm}
\be
\label{sharp}
{\dot V} + 2z V' - 4 V = (N-1) \ln (1+V') + \ln(1+V' + 2 zV'')\,,
\ee
where prime is differentiation with respect to $z$ and dot is differentiation with respect to $t$. If instead one uses Litim's optimised cutoff \cite{opt1,opt3}, eqn. \eqref{add}, in the flow for the Legendre effective action (effective average action) \cite{Nicoll1977,Wetterich:1992,Morris:1993}, then one finds \cite{Litim2005}
\be
\label{optimised}
{\dot V} + 2z V' - 4V = -\frac{N-1}{1+V'} -\frac{1}{1+V'+2zV''}+N\,.
\ee
The last term comes from a constant shift in the vacuum energy $V\mapsto V+N/4$, which has no physical consequences but streamlines the arguments later. 
And if instead we use Polchinski's equation \cite{Polchinski:1983gv}, equivalent at this level to Wilson's \cite{Wilson:1973} by a change of variables \cite{Morris:1993,Bervillier2004a,Morris:2005ck}, then for a continuous class of cutoff profiles one finds after appropriate rescalings the same flow equation \cite{Ball:1994ji}:
\be
\label{Polchinski}
{\dot U} + 2yU' - 4U = -2y\!\left(U'\right)^2 +N U' + 2y U''\,.
\ee
where for later purposes we rename the potential and fields as $U(y,t)$, where $y=\Phi^a\Phi_a/2$.

The Gaussian fixed point corresponds to the solution $V=0$ ($U=0$) 
in all three cases (\ref{sharp}--\ref{Polchinski}). Linearising about the fixed point and separating variables, we find for such infinitesimally small perturbations
\be
\label{infinitesimal}
V(z,t) = \epsilon\, w(z) \, {\rm e}^{\lambda t} \qquad{\rm and}\qquad 
U(y,t) = \epsilon\, w(y) \, {\rm e}^{\lambda t}\,,
\ee
where $\epsilon$ is considered `strictly infinitesimal', in the sense that we will for the moment accept without question the validity of the linearisation step required to form these eigen-perturbations $w$ or ``scaling fields''. The scaling dimension $\lambda$ of the associated infinitesimal coupling 
\be
\label{g}
g=\epsilon \, {\rm e}^{\lambda t}\,,
\ee 
implies a physical coupling $\p g = \epsilon \, \mu^\lambda$ of mass-dimension $\lambda$, so $g = \p g/\Lambda^\lambda$.

As advertised,  one then finds the same renormalisation group eigenvalue equation in all three cases:
\be
\label{linearised}
\lambda  w + 2 z  w' - 4  w = N  w' + 2z  w''
\ee
(up to renaming $z$ to $y$ in the third case). 
This equation is Kummer's equation and the solutions divide into two sets \cite{AbramowitzStegun,OlverAsymptotics,LebedevSpecialFunctions} depending on whether the solutions have quantised eigenvalues $\lambda$ or not. 

\subsection{Perturbations and the continuum limit}\label{sec:perturbations}

Before we develop the consequences of these solutions let us recall that, to get a {\it bona fide} field theory we need to take the continuum limit which means that we need to find a solution, called the Renormalised Trajectory, that in terms of dimensionless variables, shoots out from the fixed point\footnote{in our case the Gaussian fixed point} at $t=-\infty$ (the ultraviolet end corresponding to $\Lambda=\infty$) along a relevant (or marginally relevant) direction, and survives all the way down to the infrared end at $t=+\infty$ ($\Lambda=0$), see \eg \cite{Wilson:1973,Morris:1998}.

The exact renormalisation group may be viewed as performing a partial functional integral down to the infrared cutoff scale $\Lambda$ (see \eg 
\cite{Wetterich:1992,Morris:1993,Morris:1998}). We therefore need sensible IR behaviour at the least in order to be able to complete the integration. 

On the other hand, the UV behaviour is crucially important for a continuum limit. We require that the perturbation shoots out from the fixed point, or equivalently that the perturbations vanish into the fixed point as $t\to-\infty$, so that interactions in this sense `switch off' in the high energy limit, allowing the overall ultraviolet regulator to be removed. It is for this reason that the authors of ref. \cite{HHOrig,HHReply,HH2ndpaper} focus on this crucial property. They refer to any interactions that satisfy this as ``asymptotically free''. We will however continue to use the established term ``relevant'' (reserving ``asymptotically free'' to mean marginally relevant - as is true of the QCD coupling for example, a property which will in fact play no r\^ole in the present discussions).

\subsection{Quantised and non-quantised solutions of the linearised flow equation}\label{sec:quantised}

Returning to Kummer's equation \eqref{linearised}, we recall that the quantised perturbations are polynomials of rank $n$, the generalised Laguerre polynomials:
\be
\label{quantised}
\lambda_{n} = 4-2n\,,\qquad  w =  w_n(z) = (-2)^n L^{\frac{N}{2}-1}_n(z)\,,\qquad n=0,1,2,\cdots\,.
\ee
They thus form a complete set of orthogonal functions under the generalised Laguerre weight \cite{AbramowitzStegun,OlverAsymptotics}:
\be
\label{orthogonality}
\int_0^\infty\!\!dz\, z^{\frac{N}{2}-1}\,{\rm e}^{-z}\,  w_n(z)\,  w_m(z) = 2^{2n}\frac{\Gamma(n+N/2)}{\Gamma(n+1)}\, \delta_{n,m}\,,
\ee
allowing any potential $V(z,t)$ 
to be expanded in terms of them as 
\be
\label{expansion-V}
V(z,t) = \sum_{n=0}^\infty g_{2n}(t)\,  w_n(z)\,,
\ee
providing the potential is square-integrable with the generalised Laguerre weight, \ie providing
\be 
\label{convergence-criterion}
\int_0^\infty\!\!dz\, z^{\frac{N}{2}-1}\,{\rm e}^{-z}\,  V^2
\ee 
converges. The couplings $g_{2n}(t)$ are defined by
\be
\label{g2n}
g_{2n} = 
\frac{\Gamma(n+1)}{2^{2n}\Gamma(n+N/2)} \int_0^\infty\!\!dz\, z^{\frac{N}{2}-1}\,{\rm e}^{-z}\,  w_n(z)\, V(z,t)\,,
\ee
providing these  integrals converge.
Convergence of the series in \eqref{expansion-V} is in the almost-always sense, \ie mathematically this forms a Hilbert space structure, and convergence is with respect to the norm defined by the generalised Laguerre weight, \ie 
\be 
\label{convergence}
\int_0^\infty\!\!dz\, z^{\frac{N}{2}-1}\,{\rm e}^{-z}\, \left(V(z,t) -  \sum_{n=0}^{n_{\rm max}} g_{2n}(t)\,  w_n(z) \right)^2 \to0 \qquad {\rm as} \qquad n_{\rm max}\to\infty\,.
\ee
With the $(-2)^n$ factor in \eqref{quantised},  the highest power in $ w_n(z)$ is canonically normalised to  
\be 
\label{canon}
(\phi^a\phi_a)^n/n!\,. 
\ee
The lower powers of $\phi^a\phi_a$ in $ w_n(z)$ are generated by successive tadpole corrections. The  $(\phi^a\phi_a)^n/n!$ are of course nothing but the usual perturbative interactions with the expected scaling dimensions $\lambda_n=4-2n$ for the associated couplings $g_{2n}$ at the linearised level. Thus up to an additive constant, $ w_1(z)$ is the mass term associated to $g_2= m^2/\Lambda^2$. It has scaling dimension $\lambda_1=2$, and physical `coupling' $m^2$. Neglecting vacuum energy ($ w_0=1$), this is the sole relevant direction in the $O(N)$ symmetric potential. $ w_2(z)$ starts with the $\phi^4$ interaction, and also carries additive constant and mass-term corrections. Its associated coupling, $g_4$, is marginally irrelevant (from second order perturbation theory it vanishes as $t\to+\infty$). All higher powers, \ie $ w_n(z)$, $n\ge3$, give couplings $g_{2n}$ that are irrelevant, \ie at the linearised level they also vanish as $t\to+\infty$. 

Since from this linear analysis we see that the only relevant direction is the mass term, we recover the usual picture of triviality for scalar field theory in four dimensions which is that the continuum limit is just free massive scalar field theory. The $\Lambda\to0$ limit is sometimes known as the high temperature fixed point \cite{Zumbach:1994vg,Morris:1998,Bagnuls:2000} corresponding (in units of $\Lambda$) to an infinite mass (or zero correlation length).

The non-quantised solutions to \eqref{linearised}, corresponding to any real $\lambda\ne4-2n$ ($n=0,1,2,\cdots$) are given by Kummer M functions\cite{AbramowitzStegun,OlverAsymptotics,LebedevSpecialFunctions}:\footnote{\label{MtoL}For $\lambda=4-2n$ the M functions are just the generalised Laguerre polynomial solutions given above.}
\be
\label{Kummer}
w =  w^\lambda(z)= s_\lambda\, {\rm M}\!\!\left(\frac{\lambda-4}{2},\frac{N}{2},z\right)\,.
\ee
Here $s_\lambda$ is the sign of $(\lambda-4)(\lambda-2)$. We include it in the normalisation so that $ w^\lambda(z)$ is positive for large $z$, thus providing a perturbation to the potential that is bounded below for a positive associated coupling. The M function has a Taylor series expansion around $z=0$, and is normalised to 1 at $z=0$. In other words, $ w^\lambda(0)=s_\lambda$. Asympotically \cite{LebedevSpecialFunctions}, 
\be
\label{asymptotic}
 w^\lambda(z) = C\, {\rm e}^z\, z^{-p} \left[ 1+ O(1/z) \right]\,,
\ee
where 
\be
\label{p}
p = \frac{N}2 + 2 -\frac\lambda2 \qquad{\rm and}\qquad C = \frac{\Gamma\!\left({N/2}\right)}{ \left|\Gamma\!\left( {\lambda/2-2}\right)\right|}\,.
\ee
Since these non-polynomial solutions solve the linear equation \eqref{linearised} for real $\lambda\ne4-2n$, we already know of course that they cannot be expanded using \eqref{expansion-V}.
Substituting \eqref{asymptotic} into \eqref{g2n}, we readily confirm that (for sufficiently large $n$) these integrals do not converge. These non-quantised solutions therefore live outside the Hilbert space spanned by the quantised solutions \eqref{quantised}. However we will see in sec. \ref{sec:IR} that once correctly solved for, the IR ($t$ increasing) evolution drives them back immediately into this space. 

Since for $\lambda>0$, the $\exp (\lambda t)$ term in \eqref{g} clearly vanishes as $t\to-\infty$, Halpern and Huang \cite{HHOrig,HHReply,HH2ndpaper} conclude that these interactions are ``asymptotically free'' (\ie relevant) if $\lambda>0$.\footnote{\label{SSB}They also note that $ w^\lambda(z)$ has a minimum away from the origin if $\lambda<2$, thus potentially yielding symmetry breaking potentials,  of direct applicability for $N=4$ to the Higgs sector of the Standard Model.} Obvious as this conclusion seems to be at first sight, \emph{it is not true}, as one of us has already emphasised, albeit within the tight constraints of Physical Review Letter's comment section \cite{MorrisReb}. Our comments were based on the results already found for non-perturbative fixed points in refs. \cite{Morris:1994ie,Morris:1994jc,Morris:1994ki}, and were further expanded and developed for any fixed point in ref. \cite{Morris:1996xq}. We first review these conclusions, specialised to the current context, attempting to make them both clearer and more precise. 


\section{Ultraviolet properties: (non)existence of a continuum limit}\label{sec:ultraviolet}

Before we do so, let us underline the point that the existence of a continuum of non-quantised eigen-perturbations, \emph{is common to all fixed points} found using these techniques, not just the Gaussian fixed point considered here. In less than four dimensions, the flow equations (\ref{sharp}--\ref{Polchinski}) have non-perturbative Wilson-Fisher fixed points \cite{Wilson:1971dc,Wilson:1973} which famously are known to describe the universal properties of a wide variety of statistical physics and condensed matter systems. Perturbing around these fixed points using \eqref{infinitesimal}, it is obvious that the equations (\ref{sharp}--\ref{Polchinski}) yield linear second order ordinary differential equations and therefore it is a mathematical theorem that there are solutions to these equations for \emph{any} $\lambda$, \ie that there exist a spectrum of solutions with non-quantised values of $\lambda$.\footnote{Being second order, but with a fixed singularity at $z=0$, there is a one-parameter set of solutions for each $\lambda$, and this parameter can therefore be identified with the normalisation which we are free to choose since the equation is linear. Therefore up to normalisation, there is a unique solution for every $\lambda$.
See \eg refs. \cite{DietzMorris:2013-1,Bridle:2013sra,Dietz:2016gzg} for recent applications of these arguments in very different  exact RG contexts.}
 If one wants to maintain that the non-quantised perturbations \eqref{Kummer} actually do correspond to a continuous infinity of physical couplings $g_\lambda$ in the particle physics of a scalar field, then one also has to find a convincing rationale to exclude the equivalent perturbations in all the so far experimentally realisable continuous phase transitions described by scalar field theory (including in simulations) since a continuous spectrum of  couplings $g_\lambda$ would evidently destroy the observed universality\footnote{namely the dependence in this case on only one parameter: the overall scale, which can be identified with the correlation length.} of these phase transitions. 
 
 
%
 
So what is the logical weakness in the arguments above in sec. \ref{sec:quantised}? The linearised solutions \eqref{infinitesimal} were derived by assuming that  $V\ll1$ ($U\ll 1$). But for any fixed $\epsilon$ this condition is violated for sufficiently large $z$, since the solutions \eqref{quantised} grow like $z^n$, and solutions \eqref{Kummer} grow as $\exp z$, as in \eqref{asymptotic}. Mathematically the issue is  the difference between uniform and point-wise convergence: it is not possible to find an $\epsilon$ such that the perturbation is uniformly bounded by a $\delta>0$ of our choice. Indeed we can always find an $\epsilon$ small enough that linearisation, and thus separation of variables \eqref{infinitesimal}, is valid over some domain of $z$ and $t$,
but there is \emph{no value of $\epsilon$ no matter how small}, that ensures that linearisation of the flow equations and subsequent separation of variables is a valid procedure over the full range $0\le z<\infty$ (for any $t$). 

One might be tempted to put aside this problem for the present, for example cutting off the potential at some very large $z=\phi^a\phi_a/2$ (setting $V$ to be small or zero beyond this point), or simply assuming implicitly that $z$ can be assumed somehow sufficiently small, as is effectively done in any case in perturbation theory or more generally for scaling fields. However as noted in refs. \cite{MorrisReb,HHReply,Morris:1996xq} the large $z$ dependence is crucial to the arguments \cite{HHOrig,HHReply,HH2ndpaper} in favour of \eqref{Kummer} being a new interaction, since it is only the large $z$ behaviour \eqref{asymptotic} that prevents the integrals converging in \eqref{g2n} and thus forbids $ w^\lambda(z)$ being expanded in terms of the quantised perturbations as in \eqref{expansion-V}. \TRM{\emph{We emphasise that we have thus established that precisely in the regime where these non-polynomial solutions are distinguished from all those spanned by the quantised perturbations, it is incorrect to use the linearised solution \eqref{infinitesimal} to deduce their $t$ dependence.}}
Further steps are therefore necessary before we can draw any conclusions as to the relevance of these non-quantised perturbations, and indeed rigorously, further steps are also necessary to classify the quantised perturbations.\footnote{apart from of course the vacuum energy $w_0$ which has no field dependence}

Since the flow equations are first order differential equations in $t$, we can if we wish specify the $z$ dependence of $V(z,t)$ at some  `initial point' $t=0$. (Since $t=\ln(\mu/\Lambda)$ and $\mu$ is arbitrary, this is in fact some arbitrary point on the flow.) The $t$ evolution can  then in principle be solved for uniquely by the given flow equation. \TRM{We have seen that the evolution \eqref{linearised} is incorrect for the non-polynomial interactions \eqref{Kummer}. 
Now we ask what the correct $t$ evolution is for these interactions. In order for such a question to make any sense, one must set the potential to be \eqref{Kummer} at some initial $t$, which without loss of generality we can take to be $t=0$. (Later, in particular at the end of the paper, we will discuss the  question of the $t$ evolution of general potentials that lie outside the space spanned by the quantised interactions.) Therefore} we will proceed by setting:
\be
\label{initial}
V(z,0)=\epsilon\, v(z)\,,
\ee
agreeing with \eqref{infinitesimal} at $t=0$. Clearly however we now need a separate analysis to \eqref{infinitesimal} and \eqref{linearised} for the $t$ dependence when $z$ is large enough that we have $V(z,0)\gtrsim1$. Such an analysis will therefore have to be non-perturbative.

Obviously similar comments apply to $U(y,t)$. However in this regime, the analysis for the effective average potential (Legendre effective potential), including \eqref{optimised} and also \eqref{sharp}, will differ from that for the
general cutoff profile Polchinski's equation \eqref{Polchinski}. We will exploit this later on. For now, we concentrate on the behaviour of $V(z,t)$. 

Consider first the quantised perturbations.  The linearised mass term perturbation is given by $V(z,t)= \epsilon\,  w_1(z) \exp2t$, where $ w_1(z)=2z-N$. Setting $V(z,0)=\epsilon\,  w_1(z)$, since $ w_1(z)$ is linear inhomogeneous in $z$, we see that the right hand side of the flow equations (\ref{sharp},\ref{optimised}) are actually independent of $z$. Since the left hand side is linear anyway, the linearised solution for the $z$ piece is therefore correct whatever its magnitude, so non-perturbatively $V(z,t) = 2\epsilon\, z\exp2t +c(t)$ where $c(t)$ is to be solved for. The non-linear terms on the right hand side of the flow equation are then known explicitly thus allowing the constant term $c(t)$ to be solved for exactly. $c(t)$ therefore is not given by its linearised solution $-N\epsilon\, \exp2t$. But since there is no $z$ dependence in the non-linear terms, the $z$ dependence of the linearised solution is after all valid for all values. Furthermore, $c(t)=-N\epsilon\, \exp2t$ is a good approximation for all $t\ll t_c$ where $t_c=-\half\ln(N\epsilon)$ is the large positive RG time such that $-N\epsilon\, \exp2t\sim 1$.  We therefore conclude that the mass term $V(z,t)\to0$ as $t\to-\infty$ and thus indeed has the status of a relevant direction as the linearised analysis suggested.

Note that in physical variables \eqref{dimensions}, the mass term perturbation is
\be
\label{physical-mass}
\p V(\p z,t) =2\epsilon\,\mu^2 \p z - N\epsilon\,\mu^4\re{-2t} =2\epsilon\,\mu^2 \p z - N\epsilon\,\mu^2\Lambda^2\,,
\ee
for $t\ll t_c$. Therefore in physical variables the field dependent piece is actually independent of $t$ (\ie has mean field evolution) while the constant term \emph{actually diverges} as $t\to-\infty$. As expected, we see that it is important that we use the scaled variables appropriate to the Gaussian fixed point to determine whether or not $V(z,t)$ falls into the Gaussian fixed point ($V(z,t)\to0$) as $t\to-\infty$.

For the quantised perturbations that are (marginally) irrelevant at the linearised level ($n\ge2$), we are already certain that they do not have the correct behaviour to fall into the Gaussian fixed point since even when $V\ll 1$, the $t$ dependence is such that the perturbation grows as $t$ becomes increasingly negative.\footnote{Further comments on quantised (ir)relevant perturbations about both the Gaussian fixed point and general fixed points can be found in refs. \cite{MorrisReb,Morris:1996xq,Morris:1998}. See also sec. \ref{sec:consistency}.}

Now we answer the crucial question of whether the non-quantised perturbations \eqref{Kummer} qualify as relevant perturbations when $\lambda>0$, as suggested by the linearised analysis. Setting 
\be
\label{initial-lambda}
V(z,0) = \epsilon\,  w^\lambda(z)\,,
\ee 
we cannot now write down an analytic form for the exact solution $V(z,t)$. However the right hand side of the flow equations (\ref{sharp},\ref{optimised}) cannot contribute to the $t$ dependence of the asymptotic behaviour  \eqref{asymptotic}. In fact from \eqref{asymptotic}, for large $z$ the right hand side of \eqref{sharp} can contribute at most a term of $O(z)$ and the right hand side of \eqref{optimised} is actually exponentially suppressed. Therefore the $t$ dependence of the asymptotic expansion is found by requiring that the left hand side of the flow equations vanish. Since the left hand side of the flow equations actually came about by the dimensional assignments in \eqref{dimensions} it follows that 
in physical units the potential for large $z$ then does not evolve at all (\ie again has the so-called mean field evolution):
\be
\label{no-change}
\p V(\p z, t) = \p V(\p z,0) = \epsilon\, \p  w^\lambda(\p z)\,
\ee
(using \eqref{initial-lambda}), where we have set
\be 
\label{phys-w}
\p  w^\lambda(\p z) = \mu^4 w^\lambda(\p z/\mu^2)\,,
\ee
using  \eqref{t} and \eqref{dimensions}, while in scaled variables appropriate to determining its evolution around the Gaussian fixed point, this implies\footnote{as of course also follows directly from the vanishing of the left hand side of eqns. (\ref{sharp},\ref{optimised})}
\be
\label{mean-field}
V(z,t) = \re{4t}V(z\re{-2t},0)= \epsilon\,\re{4t} w^\lambda(z\re{-2t})\,,
\ee
\ie asymptotically
\be
\label{asymptotics+t}
V(z,t) = \epsilon\,C\,  \frac{\re{(8+N-\lambda)t}}{ z^p}\, \exp{\left(
z\re{-2t}\right)}\left[ 1+ O\left(\frac{\re{2t}}{ z}\right)\right]\,.
\ee
We see that far from falling in to the Gaussian fixed point as $t\to-\infty$, in the large $z$ regime the perturbation actually diverges rapidly as $t\to-\infty$, dominated by the $\exp\!{\left(z\re{-2t}\right)}$ term. \emph{In the large field regime, the ultraviolet behaviour of the non-quantised perturbations therefore fail to behave correctly as relevant perturbations about the Gaussian fixed point: they do not generate a Renormalised Trajectory emanating from the Gaussian fixed point and thus  cannot be used to form a continuum limit governed by the Gaussian fixed point.} 

We have seen that \eqref{asymptotics+t} can be deduced from neglecting the right hand sides of the flow equations. This neglect is justified providing $V(z,t)\gg1$. Despite its simplicity, this argument is therefore inherently non-perturbative. \TRM{In fact as mentioned above, for large $z$ the right hand side of \eqref{sharp} can contribute at most a term of $O(z)$ and the right hand side of \eqref{optimised} is actually exponentially suppressed. It is important to recognise that contributions from the right hand side of the flow equation are therefore infinitely suppressed in the large $z$ regime, in the sense that they make no contribution to the $t$ dependence of the asymptotic expansion \eqref{asymptotics+t}, even if we had carried the multiplicative series corrections in $1/z$ in \eqref{asymptotics+t} to infinite order. A closely related remark is to note that the $t$ evolution given in \eqref{asymptotics+t} can thus be used at large $z$ to derive the $t$ evolution for $V^{(n)}(z,t)$, the potential differentiated with respect to $z$ to any finite order $n$.} We emphasise again that the large field properties are crucial in this discussion since it is only these that keep $V(z,t)$ outside the space of the quantised interactions, \ie prevent it from being expanded in terms of the quantised perturbations as in \eqref{expansion-V}.

At the same time we thus see that if we replace \eqref{initial-lambda} by an initial perturbation that is a  general linear combination of the non-quantised interactions:
\be 
\label{linear-combination}
V(z,0) = \epsilon\!\int\!\! d\lambda\, \rho(\lambda)\, w^\lambda(z)\,,
\ee
for some sufficiently well-behaved density factor $\rho(\lambda)$, then either this perturbation can already be re-expressed in terms of the polynomial interactions as in \eqref{expansion-V}, and thus will have RG evolution determined by them, or it lies outside this Hilbert space in which case by \eqref{convergence-criterion} its grows at large $z$ at least as fast as $z^{-N/4}\exp(z/2)$. In this latter case we see again, by trivially adapting the argument below \eqref{asymptotics+t}, that  in the large $z$ regime the perturbation will diverge away from the Gaussian fixed point as $t\to-\infty$.

\section{IR evolution and completeness of quantised interactions}\label{sec:IR}

\subsection{Couplings in the Legendre effection action}\label{sec:couplings}

Now we put aside the ultraviolet problems and turn to explore the infrared behaviour \ie the $t>0$ domain.
We can always choose some large enough $z$ that \eqref{asymptotics+t} still applies, \ie such that neglect of the right hand side of the flow equations continues to be justified. 

To be precise, let $z=z_{asy}(t)$ be such that $V(z,t)\gg1$ according to \eqref{asymptotics+t}. Inverting, we find 
\be
\label{z-asymptotic}
z_{asy} \sim \re{2t}\ln(1/\epsilon)+A
\ee 
for some large constant $A$.\footnote{See sec. \ref{sec:WilsonianCouplings} for the higher order terms.} Then for all $t<t_f$, where $t_f$ is any fixed RG time, and for all $z>z_{asy}(t_f)$, the solution $V(z,t)$ is well approximated by \eqref{asymptotics+t}.  

Using \eqref{asymptotics+t} and approximating $w_n(z)\sim 2^n z^n/n!$ by its leading power, we thus see that for large $z$ the integrand in \eqref{g2n}  has $z$ dependence
\be
\label{z-dependence}
z^{n+\frac{\lambda}{2}-3} \re{-az}\,,\quad {\rm where}\quad a=1-\re{-2t}>0\,,
\ee
and thus the integrals defining $g_{2n}(t)$ converge.
Therefore \emph{as soon as $t>0$ (no matter how small), the couplings in the expansion \eqref{expansion-V} are well defined} \cite{MorrisReb}.

Furthermore, the expression \eqref{z-dependence} has a maximum at $z=(n-3+\lambda/2)/(1-\re{-2t})$, which is an $O(1)$ value for any $t$ larger than infinitesimal,  after which the exponential decay takes over. 
Therefore large $z$ values actually make a negligible contribution to \eqref{g2n}, and the integral is dominated by the bounded region $0<z\lesssim O(1)$. Recalling the overall $\epsilon$ multiplier, we have thus established that \emph{for all finite $t>0$, the couplings $g_{2n}(t)$ are $O(\epsilon)$.}

Finally, the series converges to the exact solution $V(z,t)$, as in \eqref{convergence}, once the integral in \eqref{convergence-criterion} converges. Using \eqref{asymptotics+t}, we see that this happens as soon as $1-2\re{-2t}>0$. Therefore we have proved that \emph{the series \eqref{expansion-V} converges, and thus $V(z,t)$ falls back into the Hilbert space spanned by the quantised interactions  $w_n(z)$, for all $t>{\ln 2}/{2}$.}

%

It is then reasonable to expect that the infrared ($t\to\infty$) fate  of the non-polynomial directions is just governed by the perturbative evolution of these quantised perturbations, which as we have reviewed, leads in the infrared to the high temperature fixed point with all interactions decayed away leaving only a diverging $w_1(z)\re{2t}$ mass term.

From \eqref{z-dependence} we see that for $n<2-\lambda/2$, the integrals $g_{2n}(t)$ converge and thus are of $O(\epsilon)$, even for $t=0$. 
For the potentially-symmetry-breaking potentials with $\lambda<2$ (see footnote \ref{SSB}) this corresponds to $O(\epsilon)$ values for the running mass $g_2(t)$ and vacuum energy $g_0(t)$ for all finite $t\ge0$.

Since we cannot have $n=2-\lambda/2$  (see footnote \ref{MtoL}) the remaining possibility for $n$ is $n>2-\lambda/2$. In this case for small positive $t$ 
the integrals are dominated by the behaviour at very large $z$ since by \eqref{z-dependence} $a\approx 2t$ for small positive $t$, so that they only just converge. We can therefore compute the leading $t$ dependence of the coefficients $g_{2n}$ from the large $z$ behaviour. Thus for $n>2-\lambda/2$ and $0<t\ll1$\footnote{In the first line, we drop the $O(1)$ integral over $0<z<z_{asy}$ and insert the leading term from $ w_n(z)$. Thus we ignore additive $O(\epsilon)$ and multiplicative $O(t)$ corrections respectively. In the second line we drop multiplicative corrections of $O(tz_{asy})$ coming from the lower limit.} 
\bea
g_{2n} &\approx& \frac{\epsilon}{2^n} \,\frac{C}{ \Gamma(n+N/2)}\,   \int_{z_{asy}}^\infty\!\!\!dz\  z^{n+\lambda/2-3}\,\re{-2tz} \nonumber\\ 
\label{t>0}
&\approx&\ \epsilon\, 2^{\lambda/2-2} \frac{\Gamma\!\left({N/2}\right)}{ \Gamma(n+N/2)} \frac{\Gamma\!\left( {n+\lambda/2-2}\right)}{ \left|\Gamma\!\left( {\lambda/2-2}\right)\right|} \left( \frac{1}{4t}\right)^{n+\lambda/2-2}\,. 
\eea
Therefore although $V(z,t)$ does have an expansion in the quantised perturbations when $t>0$, and such that the corresponding couplings are infinitesimal for finite $t>0$,
all the above couplings $g_{2n}(t)$, and in particular all the irrelevant couplings, diverge as $t\to0^+$. In this sense we see that the $t=0$ point is in fact already infinitely far from the Gaussian fixed point. 
However, as $t$ increases to positive values, these couplings rapidly shrink as expected for the irrelevant couplings and as we have seen, for $t>\ln 2/2$ the series \eqref{expansion-V} is then convergent.

Indeed from \eqref{t>0}, for sufficiently small $\epsilon$, all couplings  $g_{2n}$ of increasing irrelevancy up to some maximum $n<n_{max}(\epsilon)$ already shrink to the regime $g_{2n}(t)\ll1$ before $t$ violates the bound $t\ll1$.  We might expect to have to take $\epsilon\to0$ to form a continuum limit, despite the poor behaviour in the ultraviolet. If so, since $n_{max}\propto \ln(1/\epsilon)$ increases without bound, this means that eventually all couplings shrink to infinitesimal with increasing $t$ already for $t\ll1$.



\subsection{Consistency with large field behaviour}\label{sec:consistency}

We have seen that if the potential is set equal to an infinitesimal amount of non-polynomial direction at RG time $t=0$, \cf \eqref{initial-lambda}, then on IR evolution it immediately can be re-expressed in terms of quantised couplings which furthermore are $O(\epsilon)$, for all finite $t>0$, and such that from $t>\ln2/2$ the potential is fully inside this Hilbert space in the sense that the series then converges to the potential. This in turn gives us confidence that the universality class is just the usual one controlled by small initial values of these couplings (namely in four dimensions non-interacting scalar field theory). At first sight this picture seems deeply at variance with the fact that the large field behaviour is fixed to be mean-field, \viz \eqref{mean-field}, meaning that in physical variables the large field dependence \eqref{no-change} remains that of the original non-polynomial perturbation \eqref{phys-w} and does not actually depend on $t$ at all. 
Actually, these two pictures are consistent with each other as can be seen by the following observations. 

We have already seen from eqn. \eqref{physical-mass} that the RG properties are only manifest in scaled variables.  This is why we do not see  in physical variables that the non-polynomial potential falls back into the Hilbert space spanned by the quantised directions.  Instead in physical variables the Hilbert space  is growing to accommodate the non-polynomial interaction. This follows from \eqref{dimensions} which maps the explicit exponential in the measure in \eqref{g2n} to $\exp(-\p z/\Lambda^2)$, and which thus overcomes the $\exp(\p z/\mu^2)$  present in $\p w^\lambda$, as soon as $\Lambda<\mu$. (The latter exponential is evident from \eqref{phys-w} and \eqref{asymptotic}.)

Working again with scaled variables, we note that since the quantised couplings are $O(\epsilon)$, we expect that for finite $t$ they behave independently with the linearised $t$ dependence  of form given in \eqref{g}. Then from \eqref{quantised}, \eqref{canon} and \eqref{z}, we have for large $z$,
\be 
\label{quantised-mean-field}
g_{2n}(t)\, w_n(z) \sim g_{2n}(0)\, {\rm e}^{(4-2n) t} (2z)^n/n! \sim g_{2n}(0)\, {\rm e}^{4t} \,w_n\!\left(z\re{-2t}\right)\,,
\ee
where $g_{2n}(0)\sim\epsilon$ is the $t=0$ coefficient in  \eqref{expansion-V}. We see that for the quantised interactions the $t$-dependence of their renormalised coupling can at large $z$   be instead attributed to mean-field evolution as in \eqref{mean-field}. (This is just the RG argument for accepting these as renormalised couplings  \cite{MorrisReb,Morris:1996xq,Morris:1998}, run in reverse.) It is then natural to expect that the full sum \eqref{expansion-V} also satisfies mean-field evolution for large $z$. 

Notice that this result is general and holds for any infinitesimal perturbation that can be expanded as in \eqref{expansion-V} \ie such that the integrals in \eqref{g2n} converge.
We see that although the couplings $g_{2n}(t)$ are (marginally) irrelevant for all $n>1$, meaning that their influence should die away as $t$ increases (and completely as $t\to+\infty$), this universality property is true \emph{only for finite field. For large $\p z$ none of the quantised interactions die away and thus they preserve the large $\p z$ non-universal form of the perturbation}, since mean-field evolution is nothing more than the statement that the potential is frozen in form in physical variables, \cf \eqref{no-change}.


\subsection{Couplings in the Wilsonian effective action}
\label{sec:WilsonianCouplings}

To further address these issues, we now use the fact that the flows \eqref{optimised} and \eqref{Polchinski} are equivalent under an exact duality \cite{Morris:2005ck}. Using this map we can turn the exact solution of the flow  \eqref{optimised} for the LPA effective average potential $V(z,t)$ 
with boundary condition \eqref{initial-lambda} (and optimised cutoff), into an exact solution $U(y,t)$ of the LPA Polchinski flow equation \eqref{Polchinski} (with general cutoff profile). This $U(y,t)$ representation, which effectively just builds back in the one-particle reducible contributions missing from the one-particle irreducible $V(z,t)$ \cite{Morris:2005ck,Morris:2015oca}, has improved properties. 
This potential will also have an expansion in the quantised perturbations
\be
\label{expansion-U}
U(y,t) = \sum_{n=0}^\infty h_{2n}(t)\,  w_n(y)\,,
\ee
but the couplings $h_{2n}(t)$ in this case converge and remain $O(\epsilon)$ for all finite $t$ including for $t\le0$, and furthermore the series converges to $U(y,t)$ for all finite $t$.

In appendix \ref{app:duality}
we review the theory behind this duality. The starting point for our discussion is the generalised Legendre transform relation \cite{Morris:2005ck}:
\be
\label{duality}
U(y,t) = V(z,t) +\frac{1}{2}(\phi-\Phi)^2\,.
\ee
We have already seen that the Gaussian fixed point for the Polchinski flow \eqref{Polchinski} is $U=0$, and that infinitesimally small  perturbations  around this, namely \eqref{infinitesimal}, give the same Kummer's equation \eqref{linearised} and thus the same quantised solutions $w= w_n(y)$ as in \eqref{quantised}. As already emphasised in sec. \ref{sec:universality}, they also have the same non-quantised linearised solutions $w= w^\lambda(y)$ as in \eqref{Kummer}. However we are interested here in studying the \emph{non-perturbative} flow where the potentials are $O(1)$ or larger, generated by the large $z$ dependence of the boundary condition \eqref{initial-lambda}. Therefore we take as initial condition $U(y,0)$ as determined from
\be
\label{initial-U}
U(y,0) = \epsilon\, w^\lambda(z) +\frac{1}{2}(\phi-\Phi)^2\,,
\ee
consistently with \eqref{initial-lambda} and \eqref{duality}. The exact solution $U(y,t)$ of the flow equation \eqref{Polchinski} with this boundary condition, is then given by \eqref{duality}.

Before analysing the results of setting the boundary condition \eqref{initial-U}, we review the non-linear evolution of the mass perturbation $U(y,0)=\epsilon\,w_1(y)$ using the Wilson/Polchinski flow equation.  Its evolution follows from the analysis above \eqref{physical-mass} and the exact map \eqref{duality}, but we can just as easily solve for it directly using \eqref{Polchinski}.  This time we neglect the uninteresting vacuum energy term, so the solution takes the form $U(y,t)= 2y h_2(t)$. Substituting in \eqref{Polchinski} we find
${\dot h}_2 = 2h_2-2h_2^2$, implying 
\be
h_2(t) = {\epsilon\,\re{2t}\over1+2\epsilon\left({\rm e}^{2t}-1\right)}\,.
\ee
We see that the non-linear term on the right hand side of \eqref{Polchinski} moderates the growth of this relevant coupling in the infrared and instead of diverging like the corresponding coupling $g_2(t)=\epsilon\,\re{2t}$ of the effective average action, as analysed above \eqref{physical-mass}, it obtains a limiting value $h_2(t)\to1/2$ as $t\to\infty$. Thus 
\be
\label{HTFP}
U(y,t)=U_*(y)=y
\ee 
is in fact a fixed point of the flow, known in the literature as the high temperature fixed point. It is this `compactification' of the flow of the relevant coupling which will make it easier to see what is going on with the infrared evolution of the non-polynomial directions.

From \eqref{duality} we see that \cite{Morris:2005ck}:
\be
\Phi^a - \phi^a = \Phi^a U' = \phi^a V'\,,
\ee
Since it follows that $\phi^a$ and $\Phi^a$ point in the same
direction we deduce  \cite{DAttanasio:1997he,Morris:2005ck}
\be
\label{syz}
\sqrt{y}-\sqrt{z} = \sqrt{y}\, U' = \sqrt{z}\, V'
\ee
and hence
\be
\label{sqrts}
\sqrt{z\over y} = 1-U' = {1\over1 + V'}\,.
\ee
Thus 
\be 
\label{y(z)}
y = z\, (1+V')^2\,.
\ee
This provides us with the map from $z$ to $y$. Setting $V(z,0)=\epsilon\, w^\lambda(z)$ as in \eqref{initial-lambda}, we have that for finite $z$, \ie $z\sim O(1)$, the linearised evolution is valid for all finite $t$: $V(z,t)=\epsilon\, w^\lambda(z)\re{\lambda t}$. Thus the $V'$ term in this regime remains $O(\epsilon)$ and can be neglected in \eqref{y(z)}. Hence we see that $y=z$ for all $O(1)$ values and for all finite RG time.\footnote{Evidently $y(z)$ is then monotonic increasing, justifying the use of the Legendre transform, despite the fact (see footnote \ref{SSB}) that $w^\lambda(z)$ is initially a decreasing function and has a non-trivial minimum for $\lambda<2$.} When 
$z\to\infty$, \eqref{asymptotics+t} applies, and thus we see from the latter two equations in \eqref{sqrts} that  $U'\to1$ as $y\to\infty$. This implies that $U(y,t)$ cannot grow faster than $y$, as $y\to\infty$, for all finite RG time, and thus in fact for all RG time.

In fact we see from \eqref{sqrts} that whenever $V'\to\infty$ as $z\to\infty$, that $U'\to1$ as $y\to\infty$. This is true for any $V$ growing faster than $z$ and thus also includes all polynomial interactions \eqref{quantised} with $n>1$. We see therefore that the large field behaviour is generally under much better control in the Wilsonian picture. 

On the other hand, let us remark that the solution we are analysing is very different from the solution of the Wilson/Polchinski flow equation that one would obtain from the boundary condition $U(y,0)=\epsilon\, w^\lambda(y)$ (or by setting this to be $\epsilon\,w_n(y)$ with $n>1$). Indeed in the large $y$ region where $U(y,t)\gg1$, the right hand side of \eqref{Polchinski} cannot be neglected and thus mean-field evolution like in \eqref{asymptotics+t} does not take place. In particular the classical part itself (the first term on the right hand side of \eqref{Polchinski})  continues to contribute to the evolution of  $U(y,t)$. Furthermore, we see from \eqref{sqrts} that $z$ must vanish at the point $y_0$ where $U'(y_0,t)=1$. Therefore $z(y)$ is no longer monotonic increasing in these cases, causing the Legendre transform relation to break down.

Following \eqref{g2n}, we can compute the couplings in the expansion \eqref{expansion-U} from
\be
\label{h2n}
 h_{2n}(t) =
{\Gamma(n+1)\over2^{2n}\Gamma(n+N/2)} \int_0^\infty\!\!dy\, y^{{N\over2}-1}\,{\rm e}^{-y}\,  w_n(y)\, U(y,t)\,.
\ee
Since $U(y,t)$ grows no faster than $y$ for large $y$, these integrals always converge. Furthermore, following \eqref{convergence-criterion},the series representation \eqref{expansion-U} will always converge back to $U(y,t)$. 
In other words we have proven that \emph{the dual potential $U(y,t)$ of the non-perturbative evolution of the non-polynomial directions \eqref{initial-lambda} 
can always (\viz for all $t$) be expanded in terms the quantised perturbations such that the series \eqref{expansion-U} converges to right limit, and such that the couplings $h_{2n}(t)$ remain finite for all finite $t$.}

We have shown that for large $y$,  $U(y,t)= y$ up to corrections that grow slower than $y$. We see therefore that \emph{the large field behaviour is governed by the high temperature fixed point \eqref{HTFP} for all RG time} including the initial $t=0$. 

Now we  compute the leading correction. Again using the fact that $\phi^a$ and $\Phi^a$ point in the same direction, \eqref{duality} implies
\bea
\label{UtoV}
U(y,t) 
= (\sqrt{z}-\sqrt{y})^2 + V(z,t) &=&(\sqrt{z}-\sqrt{y})^2+ \int\!\! dz\left( \sqrt{y\over z} -1\right)  \\
&=& y-2\sqrt{zy} + \int\!\! dy\, {dz\over dy}\, \sqrt{y\over z}\,,
\eea
on using \eqref{sqrts}. The second line is in a useful form for substituting a series expansion for $z(y,t)$. Let us mention however that the first line gives, by integration by parts,  the most compact expression:
\bea
U(y,t) 
&=& (\sqrt{z}-\sqrt{y})^2-z+2\sqrt{zy} -\int\!\! {dz\over\sqrt{y}} {dy\over dz} \sqrt{z} \nonumber\\
&=& y - \int\!\!dy\,\sqrt{z\over y}\,.
\eea
Inverting \eqref{y(z)} using \eqref{asymptotics+t}, since  the $(V')^2$ term dominates, one finds $z= \re{2t} \ln (\sqrt{y}/\epsilon)+\cdots$. Therefore for both $y$ and $z$ of order of magnitude $\re{2t}\ln1/\epsilon$ or larger, one can show that
\be
\label{z(y)}
z(y,t) = \re{2t}\left\{ \ln{\sqrt{y}\over\epsilon} +\left(p-{1\over2}\right)\ln\ln{\sqrt{y}\over\epsilon} -3t-\ln C+O\left({\ln\ln{(\sqrt{y}/\epsilon)}\over\ln(\sqrt{y}/\epsilon)}\right)\right\}\,.
\ee
Thus we find that
\be
\label{asymptotics-U}
U(y,t) = y -2\re{t}\sqrt{y\,\ln{(\sqrt{y}/\epsilon)}}+O\left(\sqrt{y}\,{\ln\ln{(\sqrt{y}/\epsilon)}\over\sqrt{\ln(\sqrt{y}/\epsilon)}}\right)\,.
\ee
Despite appearances the subleading term remains of the same size as the leading term as $t$ increases, since the formula is only valid for $y\gtrsim \re{2t}\ln1/\epsilon$.  Carefully discarding the terms that are the same size or smaller than the neglected corrections $O(\cdots)$, one can verify directly that \eqref{asymptotics-U} solves the Wilson/Polchinski LPA flow equation \eqref{Polchinski}. 

Finally, we compute estimates for the couplings $h_{2n}(t)$ in the expansion \eqref{expansion-U}. From \eqref{h2n}, using the leading power for $w_n(z)$, we see they are bounded by an integral expression of form
\be
h_{2n}(t) \lesssim
{1\over2^{n}\Gamma(n+N/2)} \int_0^\infty\!\!\!dy\,\, y^{{N\over2}+n-1}\,{\rm e}^{-y}\,   U(y,t)\,.
\ee
Using \eqref{y(z)}, and \eqref{UtoV} with \eqref{syz}, we can write this as an integral over $z$:
\begin{multline}
\label{h2nz}
h_{2n}(t) \lesssim \\
{1\over2^{n}\Gamma(n+N/2)}  \int_0^\infty\!\!\!dz\,\, z^{{N\over2}+n-1}\, \left(1+V'+2zV''\right)(V+z{V'}^2)\left(1+V'\right)^{N+2n-1}\re{-z\left(1+V'\right)^2}.
\end{multline}
The integral is cut-off sharply for $z>z_1(t)$ where $z=z_1(t)$ satisfies $z{V'}^2=1$: as a consequence of \eqref{asymptotics+t},  $z\left(1+V'\right)^2$ grows exponentially as $z$ increases beyond this point. In other words, the last term in the integral vanishes as the negative exponential of an exponential for $z>z_1(t)$. Since \eqref{z(y)} in fact solves $z{V'}^2=y$, we already know that $z_1(t)=z(1,t)$, where the latter is defined in \eqref{z(y)}. 

\begin{figure}[ht]
\centering
 \includegraphics[width=0.8\textwidth]{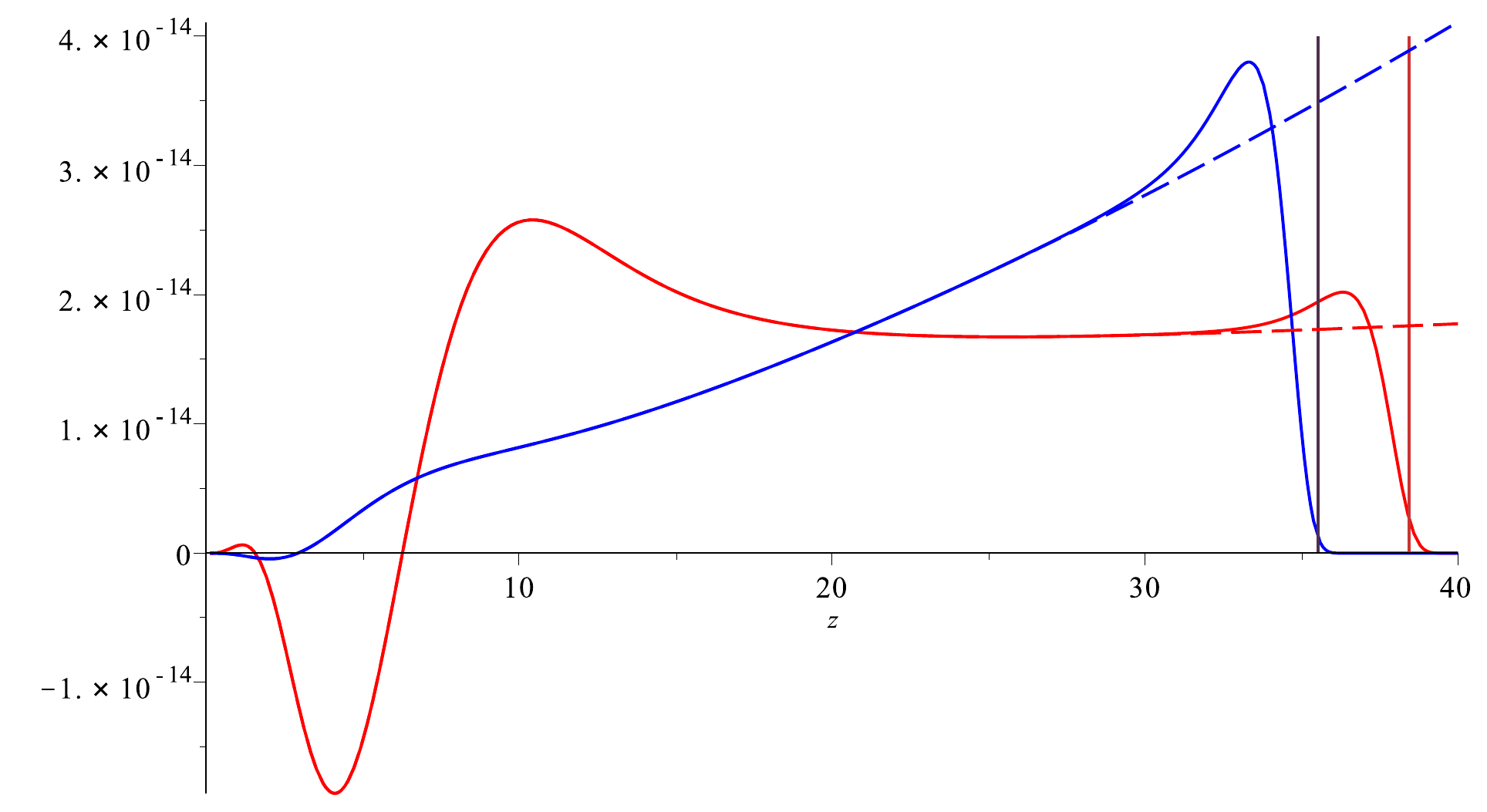} 
 \caption{The exact integrand in \eqref{h2nz} for $h_6(t)$ is plotted (together with the pre-multiplier) for the case $t=0$, as follows from \eqref{initial-lambda} and \eqref{Kummer}. We have set $N=4$, $n=3$ and $\epsilon=10^{-12}$. In red is shown the result for $\lambda=1$ (a symmetry-breaking potential) and in blue is plotted the integrand$/10$ for the case $\lambda=3$ (a symmetry preserving potential). (The small minimum in the latter case appears because the potential starts out negative.) The dotted line curves show the equivalent integrands for $g_{6}(t)$ as it appears in \eqref{integrand-o1}. The integrands only deviate once $z$ approaches $z_1$. Shown in orange and violet are the values $z=z_1(0)$ for the cases $\lambda=1,3$ respectively. The figure thus verifies that the values for $z_1$ closely approximate the effective cutoff points for the $h_{6}$ integrands.}
 \label{integrands}
\end{figure}

For $z\sim O(1)$, the integrand is the same as in \eqref{g2n}, more precisely collapses to the approximation
\be
\label{integrand-o1}
z^{{N\over2}+n-1} V(z,t) \re{-z}\,,
\ee
since in this case $V\sim O(\epsilon)$ allowing one to drop all the `correction' terms in \eqref{h2nz}. 
The balance of terms changes as $z$ approaches $z_1(t)$. From \eqref{z(y)}, we see that $z_1(t)=z(1,t)$ diverges as $z_1\sim\re{2t}\ln{1\over\epsilon}$. Thus $V' = 1/\sqrt{z_1}\sim\re{-t}/\sqrt{\ln1/\epsilon}$ and thus from \eqref{asymptotics+t}, $V\sim \re{t}/\sqrt{\ln1/\epsilon}$ and $V''\sim \re{-3t}/\sqrt{\ln1/\epsilon}$. Applying these estimates to \eqref{h2nz}, we see that in this regime the first two brackets in the integrand are now dominated by $2zV''$ and $zV'^2$ respectively, but it is the $\re{-z}\sim\epsilon^{\exp2t}$ that makes the most significant contribution.

The situation is illustrated in fig.~\ref{integrands}. Putting all these observations together we see that for finite $t>0$ only the $z\sim O(1)$ regime contributes at leading order and thus the leading estimates for the couplings in the two pictures agree $h_{2n}(t)=g_{2n}(t)\sim O(\epsilon)$. For finite $t<0$, the $z\sim z_1$ regime dominates and thus one finds $h_{2n}(t)\sim \epsilon^{\exp2t}$ to leading order. Both of these estimates receive multiplicative corrections of form $(\ln1/\epsilon)^q$ for some finite power $q$ whose precise value would require a more in-depth analysis. However already we confirm that \emph{$U(y,t)$ can be expanded as a convergent series in the quantised perturbations, as in \eqref{expansion-U}, for all $t$, and such that the couplings $h_{2n}(t)$ are infinitesimal for infinitesimal $\epsilon$ and all finite $t$.}

\section{Conclusions}\label{sec:conclusions}

In sec. \ref{sec:initial} we saw that for four-dimensional $O(N)$ scalar field theory, the LPA of different versions of the exact RG flow equation, for any sensible cutoff profile, give the same ordinary differential equation \eqref{linearised} for linearised perturbations  about the Gaussian fixed point. This is Kummer's equation and its solutions are therefore universal in this sense.
Solutions of this equation divide into polynomial perturbations \eqref{quantised}
with quantised RG eigenvalue, which are generated by the usual perturbative interactions,
 and non-polynomial perturbations \eqref{Kummer} with continuous (non-quantised) RG eigenvalue. The quantised solutions form a Hilbert space structure, allowing any potential $V(z,t)$ to be expanded in terms of them as in \eqref{expansion-V}, providing the integral \eqref{convergence-criterion} converges.
 
The only property that prevents the non-quantised perturbations $V(z,0) = \epsilon\,w^\lambda(z)$ from being expanded in terms of the quantised perturbations is their large $z$ asymptotic behaviour \eqref{asymptotic}. But as we saw in sec. \ref{sec:ultraviolet}, it is also precisely this property that prevents $V(z,t) = \epsilon\, w^\lambda(z) \, {\rm e}^{\lambda t}$ from being the correct $t$ evolution for large $z$. Instead, mean-field evolution takes over, forcing, even for $\lambda>0$, the perturbation to diverge away from the Gaussian fixed point as $t\to-\infty$. In  this way we proved that in the large field regime, the ultraviolet behaviour of the non-quantised perturbations  fail to behave correctly as relevant perturbations. 
They thus cannot generate a Renormalised Trajectory emanating from the Gaussian fixed point and therefore  cannot be used to form a continuum limit governed by this fixed point. We also saw that any linear combination of these non-quantised perturbations which is still not expandable in terms of the quantised perturbations, likewise fails to behave correctly as $t\to-\infty$.

Despite the simplicity of the argument, this $V(z,t)\gg1$ analysis is inherently non-perturbative. The same asymptotic mean-field behaviour can be established for all $z$ larger than some $z_{asy}(t)$ in the IR ($t>0$) domain. In this way in sec. \ref{sec:couplings}, we proved that 
as soon as $t>0$ (no matter how small), the couplings $g_{2n}(t)$ are well defined, while for $t>\ln2/2$ the potential $V(z,t)$ falls fully back into the Hilbert space spanned by the quantised interactions  $w_n(z)$, and can be expanded as a convergent series as in \eqref{expansion-V} \cite{Morris:1996xq}. Furthermore we proved that for all \emph{finite} $t>0$, the corresponding couplings $g_{2n}(t)$ are $O(\epsilon)$. For $n<2-\lambda/2$ the couplings are $O(\epsilon)$ even at $t=0$, while for $n>2-\lambda/2$, they diverge as a power of $t$ as $t\to0^+$. 

In this sense we see that the $t=0$ point is in fact already infinitely far from the Gaussian fixed point. With all the above properties in mind, we see why at finite $\epsilon$, the non-quantised solutions should more properly be viewed as no more special than any other finite perturbation added to the fixed point. However from  eqn. \eqref{t>0}, we also establish that in the limit as $\epsilon\to0$, all couplings shrink to infinitesimal already at arbitrarily small positive $t$. We therefore  conclude that, at least for sufficiently small $\epsilon$, the infrared ($t\to+\infty$) fate  of the non-polynomial directions is just governed by the perturbative evolution of these quantised perturbations, which (in four space-time dimensions with thus marginally irrelevant $g_4$) leads in the infrared to the high temperature fixed point with all interactions decayed away leaving only a diverging $w_1(z)\re{2t}$ mass term. 

In sec. \ref{sec:consistency} we saw why this universality conclusion is nevertheless consistent with the fact that the large field dependence remains determined by the initial perturbation, \cf \eqref{mean-field}, and indeed in physical variables fixed to be that in eqn. \eqref{no-change}. We first noted that in physical variables it is the Hilbert space that grows to accommodate this interaction. And then we noted that the universality properties in any case only hold for finite field. The self-similar evolution of the quantised couplings can instead be attributed at large field to mean-field (equivalently classical) evolution, or in physical variables no evolution at all, \cf \eqref{quantised-mean-field}. Therefore for large $\p z$ none of the quantised interactions die away and thus they preserve the large $\p z$ non-universal form of the perturbation.

Finally in sec. \ref{sec:WilsonianCouplings} we used the generalised Legendre transform relation \eqref{duality} to study the RG evolution of the corresponding exact solution $U(y,t)$ to the  Wilson/Polchinski LPA flow equation \eqref{Polchinski}. Here the high temperature fixed point is given by a genuine fixed point of the flow, and this `compactification' leads to better control of the corresponding quantised couplings $h_{2n}(t)$ in the expansion \eqref{expansion-U}. Indeed we saw that the Legendre transform relation remains valid and the large field behaviour is governed by the high temperature fixed point \eqref{HTFP}, for all finite RG time, both positive and negative. We also proved that the expansion \eqref{expansion-U} is well defined and convergent always, \ie for all $t$. Finally we found the leading dependence for the couplings in the limit $\epsilon\to0$, showing that they vanish in this limit for all finite $t$. In detail,
$h_{2n}(t)\sim \epsilon^{\exp2t}$ for finite $t<0$, while the leading behaviour $h_{2n}(t)\sim \epsilon$  agrees with that for $g_{2n}(t)$ for finite $t>0$.

As already remarked in the Introduction, and further touched on in sec. \ref{sec:ultraviolet}, although for concreteness we developed the arguments for the Gaussian fixed point and four space-time dimensions, it is clear that the arguments can be straightforwardly adapted to other space-time dimensions $d>2$ and also to the non-quantised eigen-perturbations about the Wilson-Fisher fixed points \cite{MorrisReb,Morris:1996xq,Morris:1998}.


\TRM{Finally, the main arguments can be extended to apply to the solutions of the flow equations derived at higher orders in the derivative expansion, and indeed for the exact flow equations, as we will now show.

From the above works, the $O(\partial^2)$ linearised flow equations about the Gaussian fixed point can be found, for both the potential and kinetic term interactions. In fact these can be read off essentially directly from ref. \cite{Dietz:2016gzg}  (together with expressions for the quantised polynomial solutions for both potential and kinetic terms). 

The linearised equation \eqref{linearised} for the potential itself in fact holds exactly, \ie without LPA or any other approximation, as we will now confirm. The terms on the right hand side arise from a tadpole integral \cite{Altschul:2004yq}, see also \cite{Periwal:1995hw,Dietz:2016gzg}, and as we will see, are independent of cutoff profile after rescaling variables. The  interactions satisfy the exact flow equation \eqref{Legendre2}. (From the split \eqref{split-effav} or \eqref{split-tot}, it is clear that $\Gamma$ here stands for the interactions.)
Linearisation therefore just requires to make this $\Gamma$ arbitrarily small. Therefore the exact flow equation for a perturbation about the Gaussian fixed point becomes (in unscaled variables)
\be
{\partial\over\partial\Lambda} {\Gamma}[\phi] = -{1\over2} {\rm tr}  \left[{\delta^2{\Gamma}\over\delta\phi^a\delta\phi^a}{\partial \Delta_{UV}\over\partial\Lambda}\right]\,,
\ee
where tr is a space-time trace and we have used \eqref{sum-rule} to make explicit the connection to a UV regulated tadpole diagram. If the perturbation corresponds just to a potential piece $V(z,t)$, then it is straightforward to compute the second order $\phi^a$ derivative and the above evaluates to 
\be 
\dot{V} + 2z V' -4 V = a\, (N V' +2z V'')\,,
\ee
where we used \eqref{z} and \eqref{t}, rescaled using \eqref{dimensions}, and expressed the tadpole integral as a pure (but non-universal) number:
\be 
a =\frac{1}{2\Lambda} \frac{\partial}{\partial\Lambda}\int\!\! \frac{d^4q}{(2\pi)^4} \frac{C_{UV}(q,\Lambda)}{q^2}\,.
\ee
This number can however now be absorbed by rescaling $z\mapsto a z$. After separation of variables 
\eqref{infinitesimal} we thus rederive \eqref{linearised}, this time without any approximation.

It follows immediately that the  non-polynomial scaling solutions 
\eqref{Kummer}, hold also for linearisation at the exact level. Of course none of this implies that they should lead to new relevant directions any more than they did within the LPA. On the contrary, the properties we derived, all follow from the fact that the RG evolution of (standard) scalar field theory at large field is given by mean field. This conclusion surely holds for the exact equations. Indeed it is clear from the fact that the Hessian appears as an inverse in the exact flow equations \eqref{Legendre} or \eqref{Legendre2}, that if  the Legendre effective action diverges for large field, then the right hand side of the flow equation can be neglected in this limit and thus mean-field evolution (\ie no evolution in physical variables) has to take over. The main conclusions then follow also at the exact level, in particular the non-polynomial potential perturbations diverge away from the Gaussian fixed point as $t\to-\infty$, while for $t>\ln 2/2$ they can be expanded as a convergent series in the quantised polynomial interactions.}

In fact \TRM{since} all of \TRM{our} conclusions stem from the mean field evolution \eqref{mean-field} at large field, \TRM{they} depend very little on the fact that $w^\lambda(z)$ was  a solution to Kummer's equation \eqref{Kummer}. The same conclusions would therefore follow for any initial perturbation $V(z,0)=\epsilon\, v(z)$, with large $z$ asymptotics given by \eqref{asymptotic}, even if it did not satisfy \eqref{Kummer} for some $\lambda$. Straightforward generalisations would therefore establish similar conclusions for any perturbation that grows as an exponential $z=\phi^2/2$. With some adaptation it ought to be possible to make the corresponding universality statements for initial perturbations whose growth for large field is so severe that under RG evolution they remain  forever outside the space spanned by the quantised perturbations, for example $\sim\exp(\exp\phi^2)$.

\section*{Acknowledgments}
IHB and TRM acknowledge support through an STFC studentship and  Consolidated Grant ST/J000396/1 respectively. We thank Alfio Bonanno for sharing with us his insights gained from numerical investigations of the RG flow of non-polynomial interactions.

\appendix

\section{Duality relation for effective potentials}\label{app:duality}

In this appendix we work in physical (\ie unscaled) variables although unlike in eqn. \eqref{dimensions} we will not label them explicitly as such. 
Let us recall that the flow equation for the effective average action \cite{Wetterich:1992}
\be
\label{Legendre}
{\partial\over\partial\Lambda} {\tilde\Gamma}[\phi] = {1\over2} {\rm tr}  \left[ {\cal R}\delta_{ab} + {\delta^2{\tilde\Gamma}\over\delta\phi^a\delta\phi^b}\right]^{-1} {\partial {\cal R}\over\partial\Lambda}
\ee
can trivially be alternatively written as \cite{Morris:1993,Morris:1998}
\be
\label{Legendre2}
{\partial\over\partial\Lambda} {\Gamma}[\phi] = -{1\over2} {\rm tr}  \left[ \delta_{ab} + \Delta_{IR}\cdot{\delta^2{\Gamma}\over\delta\phi^a\delta\phi^b}\right]^{-1} \Delta_{IR}^{-1} {\partial \Delta_{IR}\over\partial\Lambda}\,,
\ee
where we have split off from the effective average action a canonical kinetic term
\be
\label{split-effav}
{\tilde\Gamma}[\phi] = \Gamma[\phi] + {1\over2}\int (\partial_\mu\phi^a)^2\,,
\ee
so that the total infrared regulated Legendre effective action is now given by
\be
\label{split-tot}
\Gamma^{tot} = {1\over2}\phi^a\cdot\Delta_{IR}^{-1}\cdot\phi_a+\Gamma[\phi]\,,
\ee
where the propagator in momentum space is regulated by a multiplicative infrared regulating function $C_{IR}$:
\be
\label{multiplicative-IR}
\Delta_{IR} = {C_{IR}(q,\Lambda)/ q^2} \quad\hbox{such that}\quad q^2 C_{IR}^{-1} = q^2 +{\cal R}\,,
\ee
and $\Gamma[\phi]$ describes all the interactions (including mass terms). 

Introducing an effective ultraviolet cutoff $C_{UV}$ via the sum rule \cite{Morris:1993,Morris:1998}:
\be
\label{sum-rule}
C_{UV}+C_{IR}=1\,,
\ee
there is then an exact duality with Wilson/Polchinski's exact renormalisation group \cite{Wilson:1973,Polchinski:1983gv}:
\be
\label{pol}
{\partial S \over\partial\Lambda}={1\over2}\,
 {\delta S\over\delta\Phi^a}\cdot
{\partial\Delta_{UV}\over\partial\Lambda}\cdot {\delta
S\over\delta\Phi_a} -{1\over2}\, {\rm tr}\,
{\partial\Delta_{UV}\over\partial\Lambda}\cdot
{\delta^2S\over\delta\Phi^a\delta\Phi_a}\,,
\ee
where $S[\Phi]$ is the interaction part of the Wilsonian
effective action: 
\be
S^{tot}= \half\Phi^a.\Delta_{UV}^{-1}.\Phi_a +S[\Phi]\,,
\ee
and $\Delta_{UV}(q,\Lambda)=C_{UV}(q,\Lambda)/q^2$ is the ultraviolet regularised propagator. The flow equations \eqref{Legendre} and \eqref{pol} are therefore in fact two equivalent realizations of the same exact
RG \cite{Morris:1993}. This exact duality is implemented via the generalised Legendre transform relation \cite{Morris:1993}:
\be
\label{elegtr}
S[\Phi]=\Gamma[\phi]+\half
(\phi^a-\Phi^a)\cdot\Delta_{IR}^{-1}\cdot(\phi_a-\Phi_a)
\ee
(see also ref. \cite{Morris:2015oca}).
As shown in ref. \cite{Morris:2005ck}, remarkably this exact duality survives the Local Potential Approximation (LPA) if we employ the optimised cutoff \cite{opt1,opt3}:
 \be
\label{add}
{\cal R} = (\Lambda^2-q^2)\theta(\Lambda^2-q^2)\,.
\ee
See also refs. \cite{Litim2005,Bervillier:2007} for explorations of this. From \eqref{multiplicative-IR} we then have that
\bea
\Delta_{IR} &=&1/\Lambda^2\qquad {\rm for}\qquad
q<\Lambda\,,\nonumber\\
\label{opt}
\Delta_{IR}&=&1/q^2 \qquad\, {\rm for}\qquad q>\Lambda\,,
\eea
and thus from \eqref{sum-rule},
\bea
\Delta_{UV} &=&{1 -q^2/\Lambda^2\over q^2}\qquad {\rm for}\qquad
q<\Lambda,\nonumber\\
\Delta_{UV}&=&0 \qquad\qquad\quad\ \ \, {\rm for}\qquad q>\Lambda\,.
\eea
We see that the effective ultraviolet cutoff does indeed regulate as required inheriting its linear features and continuity from the properties of the optimised cutoff \eqref{add} \cite{Morris:2015oca}. At the LPA level effectively the fields are space-time independent. In this case the relation between interactions \eqref{elegtr}, becomes an exact relation between the effective potentials:
\be
U(\Phi) = V(\phi) +{1\over2}\Lambda^2 (\phi-\Phi)^2\,,
\ee
which in scaled variables is eqn. \eqref{duality}.

\bibliographystyle{hunsrt}
\bibliography{references}

\end{document}